\documentclass{llncs}
\usepackage{graphicx}
\usepackage{amsmath}
\usepackage{amssymb}
\usepackage{wrapfig}

\newcommand{\AbsSel}{\ensuremath{\mathsf{AbsSel}}}
\newcommand{\Atoms}{\ensuremath{\mathsf{Atoms}}}

\newcommand{\val}{\ensuremath{\mathsf{val}}}
\newcommand{\inv}{\ensuremath{\mathsf{inv}}}
\newcommand{\ite}{\ensuremath{\mathsf{ite}}}
\newcommand{\op}{\ensuremath{\mathsf{op}}}
\newcommand{\tr}{\ensuremath{\mathsf{Tr}}}
\newcommand{\plm}{\ensuremath{+_m}}
\newcommand{\isdef}{\ensuremath{\,\triangleq\,}}
\newcommand{\lsem}{\ensuremath{[\![}}
\newcommand{\rsem}{\ensuremath{]\!]}}
\newcommand{\ant}{\ensuremath{\mathsf{Ant}}}
\newcommand{\AntFail}{\ensuremath{\mathsf{AntFail}}}
\newcommand{\NoAntFail}{\ensuremath{\mathsf{NoAntFail}}}
\newcommand{\cons}{\ensuremath{\mathsf{Cons}}}
\newcommand{\constr}{\ensuremath{\mathsf{Constr}}}
\newcommand{\steword}{\ensuremath{\mathsf{STEWord}}}
\newcommand{\forte}{\ensuremath{\mathsf{Forte}}}
\newcommand{\Boolector}{\ensuremath{\mathsf{Boolector}}}
\newcommand{\true}{\ensuremath{\mathsf{true}}}
\newcommand{\false}{\ensuremath{\mathsf{false}}}
\newcommand{\arread}{\ensuremath{\mathsf{read}}}
\newcommand{\arupdate}{\ensuremath{\mathsf{update}}}
\newcommand{\mc}[1]{\ensuremath{\mathcal{{#1}}}}

\begin{document}

\title{Word-level Symbolic Trajectory Evaluation}

\author{Supratik Chakraborty\inst{1} \and 
Zurab Khasidashvili\inst{2} \and 
Carl-Johan H. Seger\inst{3} \and \\
Rajkumar Gajavelly\inst{1} \and
Tanmay Haldankar\inst{1} \and 
Dinesh Chhatani\inst{1} \and \\
Rakesh Mistry\inst{1}}
\institute{IIT Bombay, India 
\thanks{R. Gajavelly, T. Haldankar and
  D. Chhatani contributed to this work when they were in IIT Bombay.}
\and Intel IDC, Haifa, Israel \and
Intel, Portland OR, USA
}

\maketitle

\begin{abstract}
Symbolic trajectory evaluation (STE) is a model checking technique
that has been successfully used to verify industrial designs.
Existing implementations of STE, however, reason at the level of bits,
allowing signals to take values in $\{0, 1, X\}$.  This limits the
amount of abstraction that can be achieved, and presents inherent
limitations to scaling.  The main contribution of this paper is to
show how much more abstract lattices can be derived automatically from
RTL descriptions, and how a model checker for the general theory of
STE instantiated with such abstract lattices can be implemented in
practice.  This gives us the first practical word-level STE engine,
called {\steword}.  Experiments on a set of designs similar to those
used in industry show that {\steword} scales better than
word-level BMC and also bit-level STE.
\end{abstract}

\section{Introduction}

Symbolic Trajectory Evaluation (STE) is a model checking technique
that grew out of multi-valued logic simulation on the one hand, and
symbolic simulation on the other hand~\cite{BryantSeger90}.  Among
various formal verification techniques in use today, STE comes closest
to functional simulation and is among the most successful formal
verifiation techniques used in the industry.  In STE, specifications
take the form of symbolic trajectory formulas that mix Boolean
expressions and the temporal next-time operator.  The Boolean
expressions provide a convenient means of describing different
operating conditions in a circuit in a compact form.  By allowing only
the most elementary of temporal operators, the class of properties
that can be expressed is fairly restricted as compared to other
temporal logics (see~\cite{Emerson95} for a nice survey).
Nonetheless, experience has shown that many important aspects of
synchronous digital systems at various levels of abstraction can be
captured using this restricted logic.  For example, it is quite
adequate for expressing many of the subtleties of system operation,
including clocking schemas, pipelining control, as well as complex
data
computations~\cite{SegerJOMABS05,KumarGuptaGhughal12,KaivolaEtAl09}.

In return for the restricted expressiveness of STE specifications, the
STE model checking algorithm provides siginificant computational
efficiency.  As a result, STE can be applied to much larger designs
than any other model checking technique.  For example, STE is
routinely used in the industry today to carry out complete formal
input-output verification of designs with several hundred thousand
latches \cite{KumarGuptaGhughal12,KaivolaEtAl09}.  Unfortunately, this
still falls short of providing an automated technique for formally
verifying modern system-on-chip designs, and there is clearly a need
to scale up the capacity of STE even further.

The first approach that was pursued in this direction was structural
decomposition.  In this approach, the user must break down a
verification task into smaller sub-tasks, each involving a distinct STE
run.  After this, a deductive system can be used to reason about the
collections of STE runs and verify that they together imply the
desired property of the overall design~\cite{JonesOSAM01}.
In theory, structural decomposition allows verification of arbitrarily
complex designs.  However, in practice, the difficulty and tedium of
breaking down a property into small enough sub-properties that can be
verified with an STE engine limits the usefulness of this approach
significantly.  In addition, managing the structural decomposition in
the face of rapidly changing RTL limits the applicability of
structural decomposition even further.

A different approach to increase the scale of designs that can be
verified is to use aggressive abstraction beyond what is provided
automatically by current STE implementations.  If we ensure that our
abstract model satisfies the requirements of the general theory of
STE, then a property that is verified on the abstract model holds on
the original model as well.  Although the general theory of STE allows
a very general circuit model~\cite{SegerBryant95}, all STE
implementations so far have used a three-valued circuit model.  Thus,
every bit-level signal is allowed to have one of three values: $0$,
$1$ or $X$, where $X$ represents ``either $0$ or $1$''.  This limits
the amount of abstraction that can be achieved.  The main contribution
of this paper is to show how much more abstract lattices can be
derived automatically from RTL descriptions, and how the general
theory of STE can be instantiated with this lattice to give a
practical word-level STE engine that provides significant gains in
capacity and efficiency on a set of benchmarks.

Operationally, word-level STE bears similarities with word-level
bounded model checking (BMC).  However, there are important
differences, the most significant one being the use of $X$-based
abstractions on slices of words, called \emph{atoms}, in word-level
STE.  This allows a wide range of abstraction possibilities, including
a combination of user-specified and automatic abstractions -- often a
necessity for complex verification tasks.  Our preliminary
experimental results indicate that by carefully using $X$-based
abstractions in word-level STE, it is indeed possible to strike a good
balance between accuracy (cautious propagation of $X$) and performance
(liberal propagation of $X$).

The remainder of the paper is organized as follows.  We discuss how
words in an RTL design can be split into atoms in
Section~\ref{sec:atomization}.  Atoms form the basis of abstracting
groups of bits.  In Section~\ref{sec:lattice}, we elaborate on the
lattice of values that this abstraction generates, and
Section~\ref{sec:encoding} presents a new way of encoding values of
atoms in this lattice.  We also discuss how to symbolically simulate
RTL operators and compute least upper bounds using this encoding.
Section~\ref{sec:wste} presents an instantiation of the general theory
of STE using the above lattice, and discusses an implementation.
Experimental results on a set of RTL benchmarks are presented in
Section~\ref{sec:experiments}, and we conclude in
Section~\ref{sec:conclusion}.

\section{Atomizing words}\label{sec:atomization}
\begin{wrapfigure}[8]{R}{0.2\textwidth}
\begin{center}
\vspace*{-0.4in}
\includegraphics[scale=0.3]{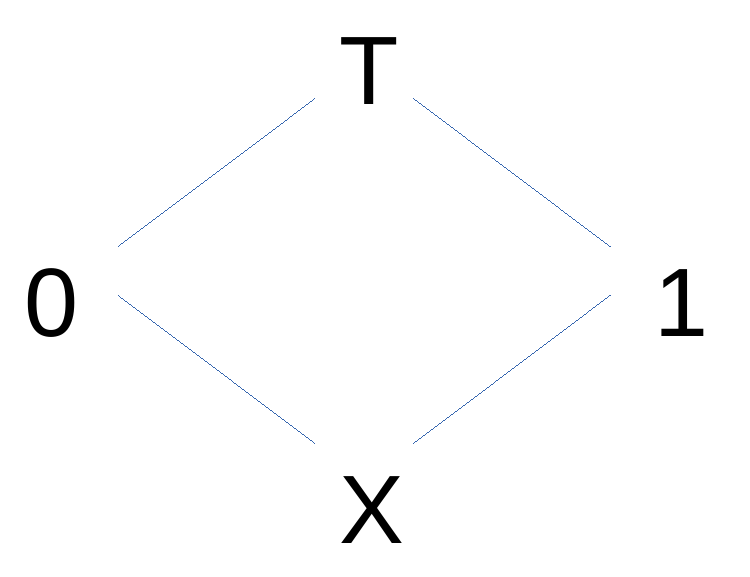}
\caption{\label{ternary-lattice}Ternary lattice}
\end{center}
\end{wrapfigure}

In bit-level STE~\cite{BryantSeger90,SegerJOMABS05}, every variable is
allowed to take values from $\{0, 1, X\}$, where $X$ denotes ``either
$0$ or $1$''.  The ordering of information in the values $0$, $1$ and
$X$ is shown in the lattice in Fig.~\ref{ternary-lattice}, where a
value lower in the order has ``less information'' than one higher up
in the order.  The element $\top$ is added to complete the lattice,
and represents an unachievable over-constrained value.  Tools that
implement bit-level STE usually use dual-rail encoding to reason about
ternary values of variables.  In dual-rail encoding, every bit-level
variable $v$ is encoded using two binary variables $v_0$ and $v_1$.
Intuitively, $v_i$ indicates whether $v$ can take the value $i$, for
$i$ in $\{0, 1\}$.  Thus, $0$, $1$ and $X$ are encoded by the
valuations $(1, 0)$, $(0, 1)$ and $(1, 1)$, respectively, of $(v_0,
v_1)$.  By convention, $(v_0, v_1) = (0, 0)$ denotes $\top$.  An
undesired consequence of dual-rail encoding is the doubling of binary
variables in the encoded system.  This can pose serious scalability
issues when verifying designs with wide datapaths, large memories,
etc.  Attempts to scale STE to large designs must therefore raise the
level of abstraction beyond that of individual bits.

In principle, one could go to the other extreme, and run STE at the level
of words as defined in the RTL design.  This requires defining a
lattice of values of words, and instantiating the general theory of
STE~\cite{SegerBryant95} with this lattice.  The difficulty with this
approach lies in implementing it in practice.  The lattice of values
of an $m$-bit word, where each bit in the word can take values in
$\{0, 1, X\}$, is of size at least $3^m$.  Symbolically representing
values from such a large lattice and reasoning about them is
likely to incur overheads similar to that incurred in bit-level STE.
Therefore, STE at the level of words (as defined in the RTL design)
does not appear to be a practical proposition for scaling.

The idea of splitting words into sub-words for the purpose of
simplifying analysis is not new (see e.g.~\cite{Joh01}).  An
aggressive approach to splitting (an extreme example being
bit-blasting) can lead to proliferation of narrow sub-words, making
our technique vulnerable to the same scalability problems that arise
with dual-rail encoding.  Therefore, we adopt a more controlled
approach to splitting.  Specifically, we wish to split words in such a
way that we can speak of an entire sub-word having the value $X$
without having to worry about which individual bits in the sub-word
have the value $X$.  Towards this end, we partition every word in an
RTL design into sub-words, which we henceforth call \emph{atoms}, such
that every RTL statement (except a few discussed later) that reads or
updates a word either does so for all bits in an atom, or for no bit
in an atom.  In other words, no RTL statement (except the few
discussed at the end of this section) reads or updates an atom partially.  

\subsubsection{Some details of atomization}
To formalize the notion of atoms, let $w$ be a word of width $m$ in an
RTL design $C$.  Let $0$ denote the least significant bit position and
$m-1$ denote the most significant bit position of $w$.  For integer
constants $p$, $q$ such that $0 \le p \le q \le m-1$, we say that the
sub-word of $w$ from bit position $p$ to $q$ is a \emph{slice} of $w$,
and denote it by $w[q:p]$.  let ${\AbsSel}(w, q, p)$ be an abstract
selection operator that either reads or writes the slice $w[q:p]$.
Concrete instances of $\mathsf{AbsSel}$ are commonly used in RTL
designs, e.g. in the System-Verilog statement {\tt c[4:1] = a[10:7] +
  b[5:2]}. We say that ${\AbsSel}(w, q, p)$ \emph{induces an
  atomization} of $w$, as shown in Table II,
where
${\Atoms}_w$ denotes the set of atoms into which $w$ is partitioned.
\begin{table}
\begin{center}
\begin{tabular}{|c|c|}
\hline
{\bfseries Condition} & {\bfseries ${\Atoms}_w$} \\
\hline
$q < m-1$ and $p > 0$ & $\{w[m-1:q+1], w[q:p], w[p-1:0]\}$ \\
\hline
$q < m-1$ and $p = 0$ & $\{w[m-1:q+1], w[q:0]\}$ \\
\hline
$q = m-1$ and $p > 0$ & $\{w[m-1:p], w[p-1:0]\}$\\
\hline
$q = m-1$ and $p = 0$ & $\{w[m-1:0]\}$\\
\hline
\end{tabular}
\caption{Computing atoms induced by ${\AbsSel}(w, q, p)$}
\end{center}
\end{table}\label{atoms}

Given atomizations ${\Atoms}^{(1)}_w$ and ${\Atoms}^{(2)}_w$, we
define their \emph{coarsest refinement} to be the atomization in which
$w[m_1:m_1]$ and $w[m_2:m_2]$ belong to the same atom iff they belong
to the same atom in both ${\Atoms}^{(1)}_w$ and
${\Atoms}^{(2)}_w$.  For every word $w[m-1:0]$ in the RTL
design, we maintain a working set, $\mathsf{WSetAtoms}_w$, of atoms.
Initially, $\mathsf{WSetAtoms}_w$ is initialized to $\{w[m-1:0]\}$.
For every concrete instance of ${\AbsSel}$ applied on $w$ in an RTL
statement, we compute ${\Atoms}_w$ using Table II, 
and
determine the coarsest refinement of ${\Atoms}_w$ and
$\mathsf{WSetAtoms}_w$.  The working set $\mathsf{WSetAtoms}_w$ is
then updated to the coarsest refinement thus computed.  The above
process is then repeated for every RTL statement in the design.

The above discussion leads to a fairly straightforward algorithm for
identifying atoms in an RTL design.  
We illustrate this on a simple example below.
\begin{wrapfigure}[16]{R}{0.5\textwidth}
\vspace*{-0.4in}
\begin{center}
\includegraphics[scale=0.4]{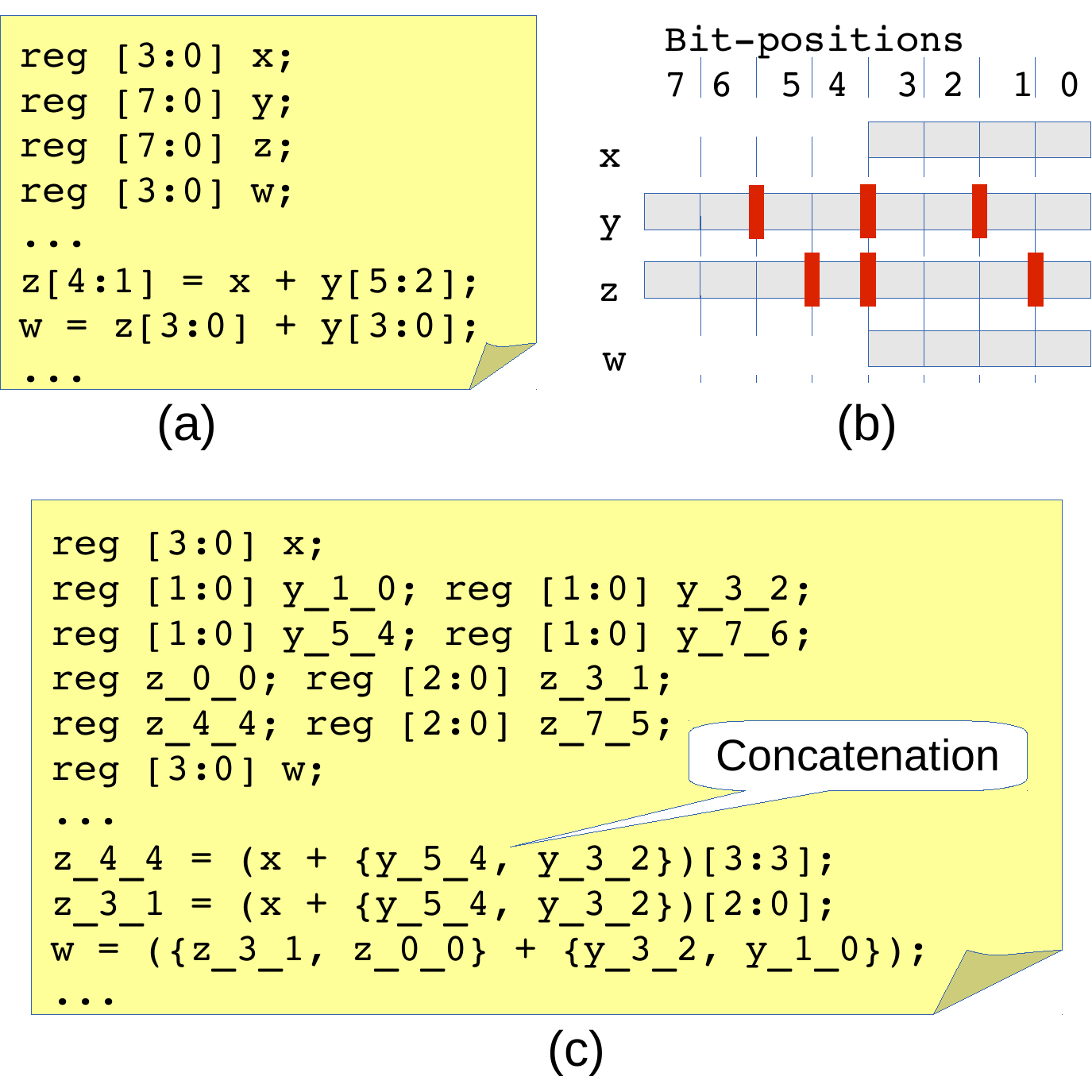}
\end{center}
\vspace*{-0.2in}
\caption{\label{atomization}Illustrating atomization}
\end{wrapfigure}
Fig.~\ref{atomization}(a) shows a System-Verilog code fragment, and
Fig.~\ref{atomization}(b) shows an atomization of words, where the
solid vertical bars represent the boundaries of atoms.  Note that
every System-Verilog statement in Fig.~\ref{atomization}(a) either
reads or writes all bits in an atom, or no bit in an atom.  Since we
wish to reason at the granularity of atoms, we must interpret
word-level reads and writes in terms of the corresponding atom-level
reads and writes.  This can be done either by modifying the RTL, or by
taking appropriate care when symbolically simulating the RTL.  For
simplicity of presentation, we show in Fig.~\ref{atomization}(c) how
the code fragment in Fig.~\ref{atomization}(b) would appear if we were
to use only the atoms identified in Fig.~\ref{atomization}(b).  Note
that no statement in the modified RTL updates or reads a slice of an
atom.  However, a statement may be required to read a slice of the
result obtained by applying an RTL operator to atoms (see, for example,
Fig.~\ref{atomization}(c) where we read a slice of the result obtained by
adding concatenated atoms). In our implementation, we do not modify the RTL.
Instead, we symbolically simulate the original RTL, but generate the
expressions for various atoms that would result from simulating the
modified RTL.

Once the boundaries of all atoms are determined, we choose to
disregard values of atoms in which some bits are set to $X$, and the
others are set to $0$ or $1$.  This choice is justified since all bits
in an atom are  read or written together.  Thus, either all bits
in an atom are considered to have values in $\{0, 1\}$, or all of them
are considered to have the value $X$.  This implies that values of an
$m$-bit atom can be encoded using $m+1$ bits, instead of using $2m$
bits as in dual-rail encoding.  Specifically, we can associate an
additional ``invalid'' bit with every $m$-bit atom.  Whenever the
``invalid'' bit is set, all bits in the atom are assumed to have the
value $X$.  Otherwise, all bits are assumed to have values in $\{0,
1\}$.  We show later in Sections~\ref{subsec:sim-val} and
\ref{subsec:sim-inv} how the value and invalid bit of an atom can be
recursively computed from the values and invalid bits of the atoms on
which it depends.

Memories and arrays in an RTL design are usually indexed by variables
instead of by constants.  This makes it difficult to atomize memories
and arrays statically, and we do not atomize them.  Similarly, if a
design has a logical shift operation, where the amount of shift is
specified by a variable, it is difficult to statically identify
subwords that are not split by the shift operation.  We ignore all
such RTL operations during atomizaion, and instead use extensional
arrays~\cite{StumpBarrettDill01} to model and reason about them.
Section~\ref{subsec:sim-inv} discusses the modeling of
memory/array reads and writes in this manner.  

\section{Lattice of atom values}\label{sec:lattice}

Recall that the primary motivation for atomizing words is to identify
the right granularity at which an entire sub-word (atom) can be
assigned the value $X$ without worrying about which bits in the
sub-word have the value $X$.  Therefore, an $m$-bit atom $a$ takes
values from the set $\{\overbrace{0\cdots 00}^\text{m bits}, \;\ldots\;
\overbrace{1\cdots 11}^\text{m bits}, \mathbf{X}\}$, where
$\mathbf{X}$ is a single abstract value that denotes an assignment of
$X$ to at least one bit of $a$.
Note the conspicuous absence of values like $0X1\cdots 0$ in the above
set.
\begin{figure}[htbp]
\begin{center}
\hspace{-0.25in}\includegraphics[scale=0.25]{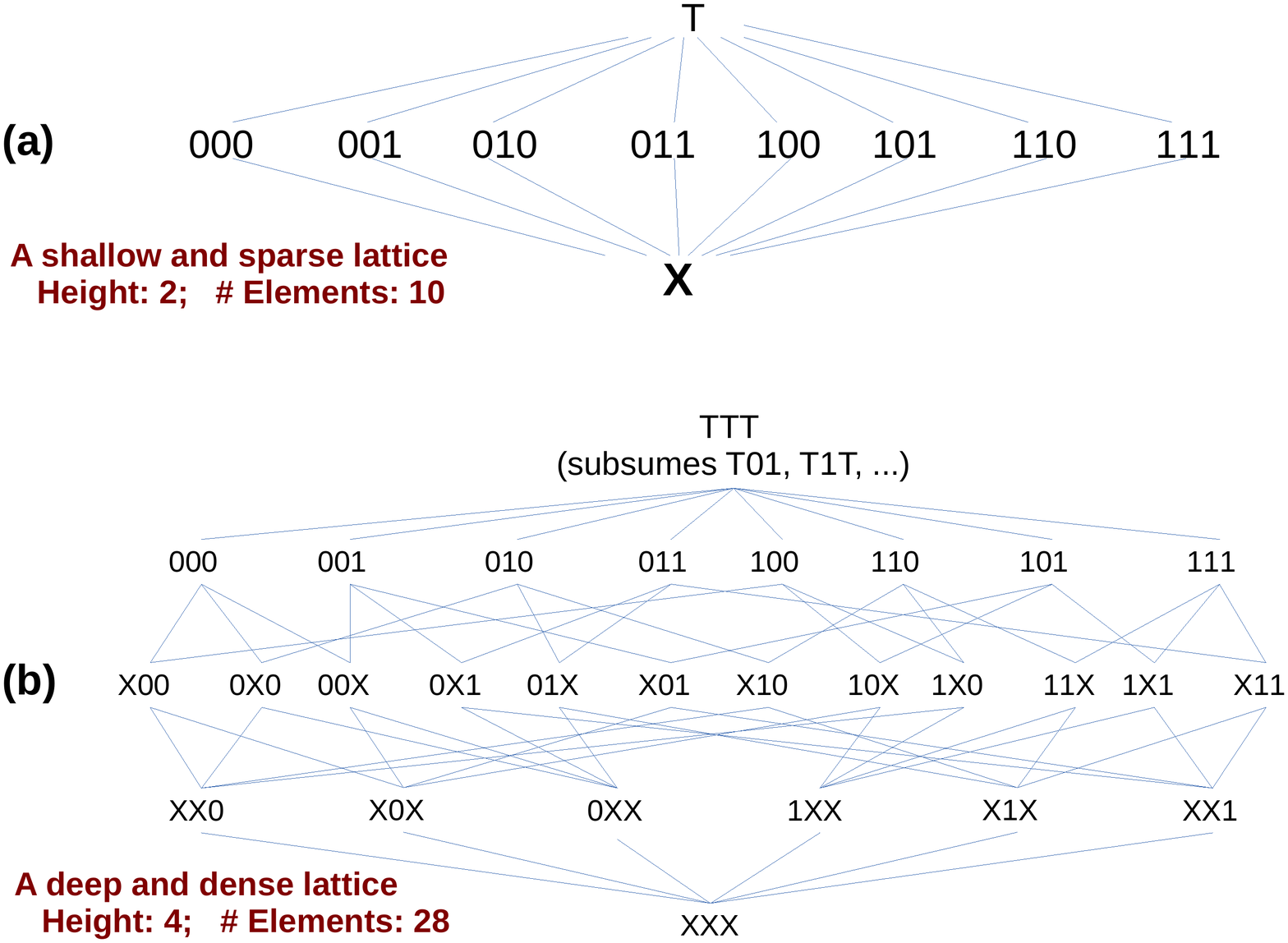}
\caption{\label{lattice-comparison} Atom-level and bit-level lattices}
\end{center}
\end{figure}
Fig.~\ref{lattice-comparison}(a) shows the lattice of values for a
$3$-bit atom, ordered by information content.  The $\top$ element is
added to complete the lattice, and represents an unachievable
over-constrained value.  Fig.~\ref{lattice-comparison}(b) shows the
lattice of values of the same atom if we allow each bit to take values
in $\{0, 1, X\}$.  Clearly, the lattice in
Fig.~\ref{lattice-comparison}(a) is shallower and sparser than that
in Fig.~\ref{lattice-comparison}(b).

Consider an $m$-bit word $w$ that has been partitioned into
non-overlapping atoms of widths $m_1, \ldots m_r$, where $\sum_{j=1}^r
m_j = m$.  The lattice of values of $w$ is given by the product of $r$
lattices, each corresponding to the values of an atom of $w$.  For
convenience of representation, we simplify the product lattice by
collapsing all values that have at least one atom set to $\top$ (and
therefore represent unachievable over-constrained values), to a single
$\top$ element.
It can be verified that the height of the product lattice (after the
above simplification) is given by $r+1$, the total number of elements
in it is given by $\prod_{j=1}^m \big(2^{m_j} + 1\big) + 1$ and the
number of elements at level $i$ from the bottom is given by
$\binom{m}{i} \prod_{j=1}^i 2^{m_j}$, where $0 < i \le r$.  It is not
hard to see from these expressions that atomization using few wide atoms (i.e.,
small values of $r$ and large values of $m_j$) gives shallow and
sparse lattices compared to atomization using many narrow atoms (i.e.,
large values of $r$ and small values of $m_j$).  The special case of a
bit-blasted lattice (see Fig.~\ref{lattice-comparison}(b)) is obtained
when $r = m$ and $m_j = 1$ for every $j \in \{1, \ldots m\}$.

Using a sparse lattice is advantageous in symbolic reasoning since we
need to encode a small set of values.  Using a shallow lattice helps
in converging fast when computing least upper bounds -- an operation
that is crucially needed when performing symbolic trajectory
evaluation.  However, making the lattice of values sparse and shallow
comes at the cost of losing precision of reasoning.  By atomizing
words based on their actual usage in an RTL design, and by abstracting
values of atoms wherein some bits are set to $X$ and the others are
set to $0$ or $1$, we strike a balance between depth and density of
the lattice of values on one hand, and precision of reasoning on the
other.

\section{Symbolic simulation with invalid-bit encoding}\label{sec:encoding}
As mentioned earlier, an $m$-bit atom can be encoded with $m+1$ bits
by associating an ``invalid bit'' with the atom.  For notational
convenience, we use ${\val}(a)$ to denote the value of the $m$ bits
constituting atom $a$, and ${\inv}(a)$ to denote the value of its
invalid bit.  Thus, an $m$-bit atom $a$ is encoded as a pair
$({\val}(a), {\inv}(a))$, where ${\val}(a)$ is a bit-vector of width
$m$, and ${\inv}(a)$ is of $\mathsf{Boolean}$ type.  Given
$({\val}(a), {\inv}(a))$, the value of $a$ is given by
${\ite}({\inv}(a), \mathbf{X}, {\val}(a))$, where ``${\ite}$'' denotes
the usual ``if-then-else'' operator.  For clarity of exposition, we
call this encoding ``invalid-bit encoding''.  Note that invalid-bit
encoding differs from dual-rail encoding even when $m=1$.
Specifically, if a $1$-bit atom $a$ has the value $X$, we can use
either $(0, \true)$ or $(1, \true)$ for $({\val}(a), {\inv}(a))$ in
invalid-bit encoding.  In contrast, there is a single value, namely
$(a_0, a_1) = (1,1)$, that encodes the value $X$ of $a$ in dual-rail
encoding.  We will see in Section~\ref{subsec:sim-inv} how this degree
of freedom in invalid-bit encoding of $X$ can be exploited to simplify
the symbolic simulation of word-level operations on
invalid-bit-encoded operands, and also to simplify the computation of
least upper bounds.

Symbolic simulation is a key component of symbolic trajectory
evaluation.  In order to symbolically simulate an RTL design in which
every atom is invalid-bit encoded, we must first determine the
semantics of word-level RTL operators with respect to invalid-bit
encoding.  Towards this end, we describe below a generic technique for
computing the value component of the invalid-bit encoding of the
result of applying a word-level RTL operator.  Subsequently, we
discuss how the invalid-bit component of the encoding is computed.

\subsection{Symbolically simulating values}\label{subsec:sim-val}
Let ${\op}$ be a word-level RTL operator of arity $k$, and let $res$
be the result of applying ${\op}$ on $v_1, v_2, \ldots v_k$, i.e.,
$res = {\op}(v_1, v_2, \ldots v_k)$.  For each $i$ in $\{1, \ldots
k\}$, suppose the bit-width of operand $v_i$ is $m_i$, and suppose the
bit-width of $res$ is $m_{res}$.  We assume that each operand is
invalid-bit encoded, and we are interested in computing the
invalid-bit encoding of a specified slice of the result, say
$res[q:p]$, where $0 \le p \le q \le m_{res} - 1$.  
Let $\langle {\op} \rangle: \{0, 1\}^{m_1} \times \cdots
\times \{0, 1\}^{m_k} \rightarrow \{0, 1\}^{m_{res}}$ denote the RTL
semantics of ${\op}$.  For example, if ${\op}$ denotes $32$-bit unsigned
addition, then $\langle {\op} \rangle$ is the function that takes two
$32$-bit operands and returns their $32$-bit unsigned sum.  The following lemma
states that ${\val}(res[q:p])$ can be computed if we know $\langle {\op}
\rangle$ and ${\val}(v_i)$, for every $i \in \{1, \ldots k\}$.
Significantly, we do not need ${\inv}(v_i)$ for any $i \in \{1, \ldots
k\}$ to compute ${\val}(res[q:p])$.  
\begin{lemma}\label{correct-val}
  Let $v = \big(\langle {\op} \rangle({\val}(v_1), {\val}(v_2), \ldots
  {\val}(v_k))\big)[q:p]$.  Then ${\val}(res[q:p])$ is given by $v$,
  where $res = {\op}(v_1, v_2, \ldots v_k)$.
\end{lemma}
\begin{proof}
  By definition of invalid-bit encoding, if ${\inv}(res[q:p])$ is
  $\true$, the value of ${\val}(res[q:p])$ does not matter.  Hence, we
  focus on the case where ${\inv}(res[q:p])$ is $\false$.  By
  definition, in this case, $res[q:p]$ has a value in $\{0,
  1\}^{q-p+1}$.  If the invalid bits of all operands $v_i$ are
  $\false$, then $\big(\langle {\op} \rangle({\val}(v_1),
  {\val}(v_2),$ $\ldots$ ${\val}(v_k))\big)[q:p]$ clearly computes the
  value of ${\val}(res[q:p])$.  Otherwise, suppose ${\inv}(v_i) =
  \true$ for some $i \in \{1, \ldots k\}$.  By definition of
  invalid-bit encoding, $v_i$ can have any value in $\{0, 1\}^{m_i}$.
  However, since ${\inv}(res[q:p])$ is $\false$, it must be the case
  that ${\val}(res[q:p])$ has a well-defined value in $\{0,
  1\}^{q-p+1}$, regardless of what value $v_i$ takes in $\{0,
  1\}^{m_i}$.  Therefore, we can set the value of $v_i$ to
  ${\val}(v_i)$ without affecting the value of $res[q:p]$.  By
  repeating this argument for all $v_i$ such that ${\inv}(v_i)$ is
  $\true$, we see that $\big(\langle {\op} \rangle({\val}(v_1),
       {\val}(v_2), \ldots {\val}(v_k))\big)[q:p]$ gives
       ${\val}(res[q:p])$.
\end{proof}
Lemma~\ref{correct-val} tells us that when computing
${\val}(res[q:p])$, we can effectively assume that invalid-bit
encoding is not used.  This simplifies symbolic simulation with
invalid-bit encoding significantly.  Note that this simplification
would not have been possible had we not had the freedom to ignore
${\val}(res[q:p])$ when ${\inv}(res[q:p])$ is $\true$.

\subsection{Symbolically simulating invalid bits}\label{subsec:sim-inv}
We now turn to computing ${\inv}(res[q:p])$.  Unfortunately, computing
${\inv}(res[q:p])$ precisely is difficult and involves
operator-specific functions that are often complicated.  We therefore
choose to approximate ${\inv}(res[q:p])$ in a sound manner with
functions that are relatively easy to compute.  Specifically, we allow
${\inv}(res[q:p])$ to evaluate to $\true$ (denoting $res[q:p] =
\mathbf{X}$) even in cases where a careful calculation would have
shown that ${\op}(v_1, v_2, \ldots v_k)$ is not $\mathbf{X}$.
However, we never set ${\inv}(res[q:p])$ to {\false} if any bit in
$res[q:p]$ can take the value $X$ in a bit-blasted evaluation of
$res$.  
Striking a fine balance between the precision and computational
efficiency of the sound approximations is key to building a
practically useful symbolic simulator using invalid-bit encoding.  Our
experience indicates that simple and sound approximations of
${\inv}(res[q:p])$ can often be carefully chosen to serve our purpose.
While we have derived templates for approximating ${\inv}(res[q:p])$
for $res$ obtained by applying all word-level RTL operators that
appear in our benchmarks, we cannot present all of them in detail here
due to space constraints.  We present below a discussion of how
${\inv}(res[q:p])$ is approximated for a subset of important RTL
operators.  Importantly, we use a recursive formulation for computing
${\inv}(res[q:p])$.  This allows us to recursively compute
invalid bits of atoms obtained by applying complex sequences of
word-level operations to a base set of atoms.  

\subsubsection{Word-level addition.} Let $\plm$ denote an
$m$-bit addition operator.  Thus, if $a$ and $b$ are $m$-bit operands,
$a \plm b$ generates an $m$-bit $sum$ and a $1$-bit $carry$.  Let the
carry generated after adding the least significant $r$ bits of the
operands be denoted $carry_r$.  We discuss below how to compute
sound approximations of ${\inv}(sum[q:p])$ and
${\inv}(carry_r)$, where $0 \le p \le q \le m-1$ and $1 \le r \le m$.

It is easy to see that the value of $sum[q:p]$ is completely
determined by $a[q:p]$, $b[q:p]$ and $carry_p$.  Therefore, we can
approximate ${\inv}(sum[q:p])$ as follows:
${\inv}(sum[q:p])\! = \! {\inv}(a[q:p]) \vee {\inv}(b[q:p]) \vee {\inv}(carry_p)$

To see why the above approximation is sound, note that if
all of ${\inv}(a[q:p])$, ${\inv}(b[q:p])$ and ${\inv}(carry_p)$ are
$\false$, then $a[q:p]$, $b[q:p]$ and $carry_p$ must have
non-$\mathbf{X}$ values.  Hence, there is no uncertainty in the value
of $sum[q:p]$ and ${\inv}(sum[q:p]) = \false$.  On the other hand, if
any of ${\inv}(a[q:p]$, ${\inv}(b[q:p])$ or ${\inv}(carry_p)$ is
$\true$, there is uncertainty in the value of $sum[q:p]$.

The computation of ${\inv}(carry_p)$ (or ${\inv}(carry_r)$) is
interesting, and deserves special attention.  We identify three cases
below, and argue that ${\inv}(carry_p)$ is $\false$ in each of these
cases.  In the following, $\mathbf{0}$ denotes the $p$-bit constant
$00\cdots 0$.
\begin{enumerate}
\item If $\big({\inv}(a[p-1:0]) \vee {\inv}(b[p-1:0])\big) = \false$,
  then both ${\inv}(a[p-1:0])$ and ${\inv}(b[p-1:0])$ must be
  $\false$.  Therefore, there is no uncertainty in the values of
  either $a[p-1:0]$ or $b[p-1:0]$, and ${\inv}(carry_p) = \false$.
\item If $\big(\neg{\inv}(a[p-1:0]) \wedge ({\val}(a[p-1:0]) =
  \mathbf{0})\big)$, then the least significant $p$ bits of
  ${\val}(a)$ are all $0$.  Regardless of ${\val}(b)$, it is easy to
  see that in this case, ${\val}(carry_p) = 0$ and ${\inv}(carry_p) =
  \false$.
\item This is the symmetric counterpart of the case above, i.e.,
  $\big(\neg{\inv}(b[p-1:0]) \wedge ({\val}(b[p-1:0]) = \mathbf{0})\big)$.
\end{enumerate}
We now approximate ${\inv}(carry_p)$ by combining the conditions
corresponding to the three cases above.  In other words,
\begin{eqnarray}
 {\inv}(carry_p)  & =  & \big({\inv}(a[p-1:0]) \! \vee \! {\inv}(b[p-1:0])\big) \! \wedge \! \nonumber\\
               \!\!  & \! \!  & \! 
\big({\inv}(a[p-1:0]) \! \vee \! ({\val}(a[p-1:0]) \!  \neq  \! \mathbf{0})\big) \! \wedge \nonumber\\
              \! \!  & \!\!  & \!\big({\inv}(b[p-1:0]) \!\vee \!({\val}(b[p-1:0]) \!\neq  \! \mathbf{0})\big) \nonumber
\end{eqnarray}

\subsubsection{Word-level division.} Let $\div_m$ denote an
$m$-bit division operator; this is among the most complicated
word-level RTL operators for which we have derived an approximation of
the invalid bit.  If $a$ and $b$ are $m$-bit operands, $a \div_m b$
generates an $m$-bit quotient, say $quot$, and an $m$-bit remainder,
say $rem$.  We wish to compute ${\inv}(quot[q:p])$ and
${\inv}(rem[q:p])$, where $0 \le p \le q \le m-1$.  We assume that if
${\inv}(b)$ is $\false$, then $b \neq 0$; the case of $a \div_m b$
with $({\val}(b), {\inv}(b)) = (0, \false)$ leads to a
``divide-by-zero'' exception, and is assumed to be handled separately.

The following expressions give sound approximations for
${\inv}(quot[q:p])$ and ${\inv}(rem[q:p])$.  In these expressions, we
assume that $i$ is a non-negative integer such that $2^i \le {\val}(b)
< 2^{i+1}$.  

\begin{eqnarray}
{\inv}(quot[q:p]) & = & {\ite}({\inv}(b), \,temp_1, \, temp_2), \text{ where } \nonumber\\
temp_1            & = & {\inv}(a) \,\vee\, ({\val}(a[m-1:p]) \neq \mathbf{0}) \text{ and }\nonumber \\
temp_2            & = & {\ite}({\val}(b) = 2^i, \,temp_3, \,(i < p) \,\vee\, 
{\inv}(a[m-1:p])), \text{ where } \nonumber \\
temp_3            & = & (p+i \le m-1) \,\wedge\, 
{\inv}(a[\min(q+i,m-1):p+i])) \nonumber\\
{\inv}(rem[q:p])\! & = & \!{\inv}(b) \,\vee\, {\ite}({\val}(b) = 2^i, (i > p) \wedge 
{\inv}(a[\min(q, i-1):p]), i \ge p) \nonumber 
\end{eqnarray}
Note that the constraint $2^i \le {\val}(b) < 2^{i+1}$ in the
above formulation refers to a fresh variable $i$ that does not
appear in the RTL.  
We will see later in Section~\ref{sec:wste} that a word-level STE
problem is solved by generating a set of word-level constraints, every
satisfying assignment of which gives a counter-example to the
verification problem.  We add constraints like $2^i \le {\val}(b) <
2^{i+1}$ in the above formulation, to the set of word-level
constraints generated for an STE problem.  This ensures that every
assignment of $i$ in a counterexample satisfies the required
constraints on $i$.


To see why the above approximations for ${\inv}(quot[q:p])$ and
${\inv}(rem[q:p])$ are sound, first consider the case where
${\inv}(b) = \true$.  Since we are unsure of the value of the divisor,
not much can be said about the remainder.  So, we set
${\inv}(rem[q:p])$ to $\true$.  The situation is slightly better for
the quotient.  If we know that ${\inv}(a) = \false$, then since the
quotient of integer division is never larger than the dividend, we can
infer that $quot[q:p] = 0$ if $a[m-1:p] = 0$.  Clearly, in this case
${\inv}(quot[q:p]) = \false$.  In all other sub-cases of ${\inv}(b) =
\true$, we set ${\inv}(quot[q:p])$ to $\true$.

If ${\inv}(b) = \false$, we know that $b$ has a value in $\{0,
1\}^{m}$, but not $\mathbf{0}$.  Representing bit vectors by their
integer representations, let $i \in \{0, \ldots m-1\}$ be such that
$2^i \le {\val}(b) < 2^{i+1}$.  We consider two
sub-cases below.
\begin{itemize}
\item ${\val}(b) = 2^i :$ In this case, $a \div_m b$ effectively
  shifts $a$ right by $i$ bit positions, and the least significant $i$
  bits of $a$ forms the remainder.
  Therefore, ${\val}(quot[q:p])$ is $a[q+i:p+i]$ if $q+i \le
  m-1$, is $a[m-1:p+i]$ padded to the left with $q-m+i+1$ $0$s if $q+i
  > m-1 \le p+i$, and is $0$ if $p+i > m-1$.
  It follows that if $p+i > m-1$, then ${\val}(quot[q:p]) = {\mathbf 0}$
  and ${\inv}(quot[q:p]) = \false$.  Otherwise, ${\inv}(quot[q:p]) =
  {\inv}(a[k:p+i])$, where $k = \min(q+i, m-1)$.
it is easy to see that ${\val}(rem[q:p])$ is
$a[q:p]$ if $i > q$, is $a[i-1:p]$ padded with $q-i+1$ $0$s to the
left if $q \ge i > p$, and is $0$ if $i \le p$.  
  By similar reasoning, if $i \le p$, then
${\inv}(rem[q:p]) = \false$;  otherwise, ${\inv}(rem[q:p]) =
{\inv}(a[k:p])$, where $k = \min(q, i-1)$.
\item $2^i < {\val}(b) < 2^{i+1} :$ In this case, we show below that
  if $i \ge p$, then ${\inv}(quot[q:p])$ can be approximated
  by ${\inv}(a[m-1:p])$.  If $i < p$, then ${\inv}(rem[q:p]) =
  \false$.  In all other cases, we approximate ${\inv}(quot[q:p])$ and
  ${\inv}(rem[q:p])$ by $\true$.

  To see why the above approximations are sound, note that
  ${\val}(a)$ can be written as $a_1\cdot 2^p + a_2$, where $a_1$ and
  $a_2$ are the integer representations of $a[m-1:p]$ and $a[p-1:0]$,
  respectively.  Clearly, $0 \le a_2 < 2^p$.  Considering quotients
  and remainders on division by ${\val}(b)$, suppose $a_1 = k_1\cdot
  {\val}(b) + r_1$ and $a_2 = k_2\cdot {\val}(b) + r_2$, where $0 \le
  r_1, r_2 < {\val}(b)$ and $k_1, k_2 \ge 0$.  Suppose further that
  $2^p\cdot r_1 + r_2 = k_3\cdot {\val}(b) + r_3$, where $0 \le r_3 <
  {\val}(b)$ and $k_3 \ge 0$.  It is an easy exercise to see that the
  quotient of dividing ${\val}(a)$ by ${\val}(b)$ is $2^p\cdot k_1 +
  k_2 + k_3$, and the remainder is $r_3$.  Thus, ${\val}(quot) =
  2^p\cdot k_1 + k_2 + k_3$ and ${\val}(rem) = r_3$.  We discuss what
  happens when $i \ge p$ and $i+1 \le p$.
  \begin{itemize}
    \item If $i \ge p$, then ${\val}(b) > 2^i \ge 2^p > a_2$.  Since
      ${\val}(b) > a_2$, we have $k_2 = 0$ and $r_2 = a_2 < 2^p$.  It
      follows that $quot = 2^p\cdot k_1 + k_3$.  If $k_3 < 2^p$, then
      $quot[q:p]$ depends only on $k_1$, which in turn, depends only
      on $a[m-1:p]$ and ${\val}(b)$.  Therefore, ${\inv}(quot[q:p])$
      can be approximated by ${\inv}(a[m-1:p])$.

      We now show that $k_3$ is indeed strictly less than $2^p$.
      Since $2^p\cdot r_1 + r_2 = k_3\cdot {\val}(b) + r_3$,
      rearranging terms, we get $k_3\cdot {\val}(b) - 2^p\cdot r_1 =
      r_2 - r_3$.  If possible, let $k_3 = 2^p + d$, where $d \ge 0$.
      Substituting for $k_3$, we get $2^p\cdot({\val}(b) - r_1) +
      d\cdot {\val}(b) = r_2 - r_3$.  Since ${\val}(b) > r_1$, the
      left hand side of the above equation is at least as large as $
      2^p$, while the right hand side is at most $r_2$, which, in
      turn, is less than $ 2^p$.  This gives a contradiction, and
      therefore, $k_3 < 2^p$.
    \item If $i < p$, we have $rem = r_3 < {\val}(b) < 2^{i+1} \le 2^p$.
      Therefore, ${\val}(rem[q:p]) = 0$, and ${\inv}(rem[q:p]) = \false$.
  \end{itemize}
\end{itemize}
The above analysis yields the sound approximations for
${\inv}(quot[q:p])$ and ${\inv}(rem[q:p])$ discussed above.

\subsubsection{If-then-else statements.} Consider a conditional
assignment statement ``{\tt if (BoolExpr) then x = Exp1; else x =
  Exp2;}''.  Symbolically simulating this statement gives $x =
{\ite}(\mathsf{BoolExpr}, \mathsf{Exp1}, \mathsf{Exp2})$.  The
following gives a sound approximation of ${\inv}(x[q:p])$. 
\begin{eqnarray}
{\inv}(x[q:p]) & = & {\ite}({\inv}(\mathsf{BoolExpr}), \, temp_1, \, temp_2), \text{ where } \nonumber \\
temp_1         & = & {\inv}(\mathsf{Exp1}[q:p])\,\vee\,{\inv}(\mathsf{Exp2}[q:p])\,\vee\,
({\val}(\mathsf{Exp1}[q:p]) \ne {\val}(\mathsf{Exp2}[q:p])) \nonumber \\
temp_2         & = & {\ite}({\val}(\mathsf{BoolExpr}), \,{\inv}(\mathsf{Exp1}[q:p]), 
{\inv}(\mathsf{Exp2}[q:p])) \nonumber 
\end{eqnarray}

To see why the above approximation of ${\inv}(x[q:p])$ is sound, let
$x = {\ite}(\mathsf{BoolExpr}, \mathsf{Exp1}, \mathsf{Exp2})$, where
$\mathsf{BoolExpr}$ is a boolean expression, and $\mathsf{Exp1}$ and
$\mathsf{Exp2}$ are expressions of the same type as $x$.  To compute
${\inv}(x[q:p])$, we note that if ${\inv}(\mathsf{BoolExpr}) =
\false$, then ${\inv}(x[q:p])$ is simply
${\ite}({\val}(\mathsf{BoolExpr}), \,{\inv}(\mathsf{Exp1}[q:p]),
\,{\inv}(\mathsf{Exp2}[q:p]))$.  However, if
${\inv}(\mathsf{BoolExpr}) = \true$, then the value of ${BoolExpr}$
could be $1$ (denoting $\true$) or $0$ (denoting $\false$).
Interestingly, if both ${\inv}(\mathsf{Exp1}[q:p]$ and
${\inv}(\mathsf{Exp2}[q:p])$ are $\false$ (i.e., neither
$\mathsf{Exp1}[q:p]$ nor $\mathsf{Exp2}[q:p]$ are $\mathbf{X}$), and
if ${\val}(\mathsf{Exp1}[q:p]) = {\val}(\mathsf{Exp2}[q:p])$, then
regardless of the value of $\mathsf{BoolExpr}$, we have
${\inv}(x[q:p]) = \false$.  This is formalized in the approximation
for ${\inv}(x[q:p])$ mentioned above.

\subsubsection{Bit-wise logical operations.} Let $\neg_m$
and $\wedge_m$ denote bit-wise negation and conjunction operators
respectively, for $m$-bit words.  If $a$, $b$, $c$ and $d$ are $m$-bit
words such that $c = \neg_m a$ and $d = a \wedge_m b$, it is easy to
see that the following give sound approximations of ${\inv}(c)$ and
${\inv}(d)$.
\begin{eqnarray}
{\inv}(c[q:p]) & = & {\inv}(a[q:p]) \nonumber \\
{\inv}(d[q:p]) & = & \big({\inv}(a[q:p]) \vee {\inv}(b[q:p])\big) \,\wedge 
\big({\inv}(a[q:p]) \vee ({\val}(a[q:p]) \neq \mathbf{0})\big)\,\wedge\,\nonumber\\
& & \big({\inv}(b[q:p]) \vee ({\val}(b[q:p]) \neq \mathbf{0})\big)\nonumber 
\end{eqnarray}
The invalid bits of other bit-wise logical operators (like
disjunction, xor, nor, nand, etc.) can be obtained by first expressing
them in terms of $\neg_m$ and $\wedge_m$ and then using the above
approximations.

\subsubsection{Memory/array reads and updates.} Let
$\mathsf{A}$ be a $1$-dimenstional array, $\mathsf{i}$ be an index
expression, and $\mathsf{x}$ be a variable and $\mathsf{Exp}$ be an
expression of the base type of $A$.  On symbolically simulating the
RTL statement ``{\tt x = A[i];}'', we update the value of $\mathsf{x}$
to ${\arread}(\mathsf{A}, \mathsf{i})$, where the ${\arread}$ operator
is as in the extensional theory of arrays
(see~\cite{StumpBarrettDill01} for details).  Similarly, on
symbolically simulating the RTL statement ``{\tt A[i] = Exp}'', we
update the value of array $\mathsf{A}$ to
${\arupdate}(\mathsf{A}_{\text{orig}}, \mathsf{i}, \mathsf{Exp})$,
where $\mathsf{A}_{\text{orig}}$ is the (array-typed) expression for
$\mathsf{A}$ prior to simulating the statement, and the ${\arupdate}$
operator is as in the extensional theory of arrays. 

Since the expression for a variable or array obtained by symbolic
simulation may now have ${\arread}$ and ${\arupdate}$ operators, we
must find ways to compute sound approximations of the invalid bit for
expressions of the form ${\inv}({\arread}(\mathsf{A},
\mathsf{i})[q:p])$. Note that since $\mathsf{A}$ is an array, the
symbolic expression for $\mathsf{A}$ is either (i)
$\mathsf{A}_{\text{init}}$, i.e. the initial value of $\mathsf{A}$ at
the start of symbolic simulation, or (ii) ${\arupdate}(\mathsf{A}',
\mathsf{i}', \mathsf{Exp}')$ for some expressions $\mathsf{A}'$,
$\mathsf{i}'$ and $\mathsf{Exp}'$, where $\mathsf{A}'$ has the same
array-type as $\mathsf{A}$, $\mathsf{i}'$ has an index type, and
$\mathsf{Exp}'$ has the base type of $\mathsf{A}$.  For simplicity of
exposition, we assume that all arrays are either completely
initialized or completely uninitialized at the start of symbolic
simulation.  The invalid bit in case (i) is then easily seen to be
${\true}$ if $\mathsf{A}_{\text{init}}$ denotes an uninitialized
array, and ${\false}$ otherwise.  In case (ii), let $v$ denote
${\arread}(\mathsf{A}, \mathsf{i})$.  The invalid bit of $v[q:p]$ can
then be approximated as:
\begin{eqnarray}
{\inv}(v[q:p]) & =&   {\inv}(\mathsf{i}) \vee {\inv}(\mathsf{i}')\vee 
{\ite}\left({\val}(\mathsf{i}) = {\val}(\mathsf{i}'), {\inv}(\mathsf{Exp}'[q:p]), temp\right), \text{ where} \nonumber\\
temp & = &{\inv}({\arread}(\mathsf{A}', \mathsf{i})[q:p]).\nonumber 
\end{eqnarray}

To see why the above expression gives a sound approximation of
${\inv}(v[q:p])$, note that if either $\mathsf{i}$ or $\mathsf{i}'$ is
$\mathbf{X}$ (i.e. the corresponding invalid bit is ${\true}$), we
conservatively set ${\inv}({\arread}({\arupdate}(\mathsf{A}',
\mathsf{i}', \mathsf{Exp}'), \mathsf{i})$ to $\true$.  If neither
$\mathsf{i}$ nor $\mathsf{i}'$ is $\mathbf{X}$, there are two cases to
consider.
\begin{itemize}
\item If ${\val}(\mathsf{i}) = {\val}(\mathsf{i}')$, then
  ${\arread}({\arupdate}(\mathsf{A}', \mathsf{i}', \mathsf{Exp}'),
  \mathsf{i}) = \mathsf{Exp}'$.  Hence, the required invalid bit is
  ${\inv}(\mathsf{Exp}'[q:p])$.
\item If ${\val(\mathsf{i})} \ne {\val}(\mathsf{i}')$, then
  ${\arread}({\arupdate}(\mathsf{A}', \mathsf{i}', \mathsf{Exp}'),
  \mathsf{i}) = {\arread}(\mathsf{A}', \mathsf{i})$. Hence, the
  required invalid bit is ${\inv}({\arread}(\mathsf{A}',
  \mathsf{i})[q:p])$.
\end{itemize}
If the RTL design has multi-dimensional arrays, we simply treat them as arrays of
arrays, and apply the same reasoning as above.  For example, if
$\mathsf{B}$ is a two-dimenstional array, the RTL statement ``{\tt
  B[i][j] = Exp;}'' updates the symbolic value of array $\mathsf{B}$
to ${\arupdate}(\mathsf{B}_{\text{orig}}, \mathsf{i},
{\arupdate}({\arread}(\mathsf{B}_{\text{orig}}, \mathsf{i}),
\mathsf{j}, \mathsf{Exp}))$, where $\mathsf{B}_{\text{orig}}$ is the
symbolic expression for $\mathsf{B}$ prior to simulating the RTL statement.
Similarly, the RTL statement ``{\tt x = B[i][j];}''updates the
symbolic value of $\mathsf{x}$ to ${\arread}({\arread}(\mathsf{B}, \mathsf{i}), \mathsf{j})$.

\subsubsection{Shift operations.} We discuss below the
left-shift operation; the case of the right-shift operation can be
analyzed similarly.  A shift operation can specify either a constant
number of bit positions to shift, or a variable number of positions to
shift.  We analyze these two cases separately since shifting by a
variable number of positions does not allow us to statically identify
the operand's bit-slices of interest.  In either case, we assume that
a left shift operation pads $0$s in the least signficant shifted
positions.  Let $\ll_k$ denote a unary left-shift operator of the
first kind, where $k$ is a positive integer constant, and let $\ll$
denote a binary left-shift operator of the second kind.  Let $a, b, c,
d$ be $m$-bit words such that $b \,=\, \ll_k a$ and $c = a \ll d$.
For simplicity of presentation, we assume no wrap-around in shifting;
the case of wrap-around can be analyzed in a similar way.  The
following equations give sound approximations of ${\inv}(b[q:p])$ and
${\inv}(c[q:p])$, where $0 \le p \le q \le m-1$.
\vspace*{-0.03in}
\begin{eqnarray}
{\inv}(b[q:p]) \!\!&\!\!=\!\! & \!\! {\ite}(p \ge k, {\inv}(a[q-k:p-k]), temp), \!\!\text{ where }\nonumber \\
temp      \!\!   &\!\! =\!\! & \!\! {\ite}(q \ge k, \,{\inv}(a[q-k:0]), \,{\false}) \!\!\!\!\!\! \!\!\!\!\!\!\!\!\!\! \!\!\!\!\!\!\!\!\!\!\!\!\!\!\!\!  \label{inv-lshift-const}\\
{\inv}(c[q:p]) \!\!& \!\!= \!\! &\!\! {\inv}(a[q:0]) \,\wedge\, ({\inv}(d) \vee ({\val}(d) \leq q)) \!\!\!\!\!\!\!\!\!\!\!\!\!\!\!\!    \!\!\!\!\!\!\!\!\!\!\!\!\!\!\!\!  \label{inv-lshift-var}
\end{eqnarray}

\subsection{Computing least upper bounds}\label{subsec:lub}
Let $a = ({\val}(a), {\inv}(a))$ and $b = ({\val}(b), {\inv}(b))$ be
invalid-bit encoded elements in the lattice of values for an
$m$-bit atom.  We define $c = lub(a, b)$ as follows.
  \begin{itemize}
    \item[(a)] If $(\neg{\inv}(a) \wedge \neg{\inv}(b) \wedge ({\val}(a)
      \ne {\val}(b))$, then $c = \top$.
    \item[(b)] Otherwise, ${\inv}(c) = {\inv}(a) \wedge {\inv}(b)$ and
      ${\val}(c) = {\ite}({\inv}(a), \,{\val}(b), \,{\val}(a))$ (or
      equivalently ${\val}(c) = {\ite}({\inv}(b), \,{\val}(a), \,{\val}(b))$).
  \end{itemize}
Note the freedom in defining ${\val}(c)$ in case (b) above.  This
freedom comes from the observation that if ${\inv}(c) = \true$, the
value of ${\val}(c)$ is irrelevant.  Furthermore, if the condition in
case (a) is not satisfied and if both ${\inv}(a)$ and ${\inv}(b)$ are
$\false$, then ${\val}(b) = {\val}(c)$.  This allows us to simplify
the expression for ${\val}(c)$ on-the-fly by replacing it with
${\val}(b)$, if needed.  

\section{Word-level STE}\label{sec:wste}
In this section, we briefly review the general theory of
STE~\cite{SegerBryant95} instantiated to the lattice of values of
atoms.  An RTL design $C$ consists of inputs, outputs and internal
words.  We treat bit-level signals as $1$-bit words, and uniformly
talk of words.  Every input, output and internal word is assumed to be
atomized as described in Section~\ref{sec:atomization}.  Every atom of
bit-width $m$ takes values from the set $\{\mathbf{0} \ldots
\mathbf{2^m-1}, \mathbf{X}\}$, where constant bit-vectors have been
represented by their integer values.  The values themselves are
ordered in a lattice as discussed in Section~\ref{sec:lattice}.  Let
$\leq_m$ denote the ordering relation and $\sqcup_m$ denote the $lub$
operator in the lattice of values for an $m$-bit atom.  The lattice of
values for a word is the product of lattices corresponding to every
atom in the word.  Let $\mc{A}$ denote the collection of all
atoms in the design, and let $\mc{D}$ denote the collection
of values of all atoms in $\mc{A}$.  A state of the design
is a mapping $s: \mc{A} \rightarrow \mc{D} \cup {\top}$ such that if
$a \in \mc{A}$ is an $m$-bit atom, then $s(a)$ is a value in the set
$\{\mathbf{0}, \ldots \mathbf{2^m-1}, \mathbf{X}, \top\}$.  Let
$\mc{S}$ denote the set of all states of the design.  Clearly $\mc{S}$
forms a lattice -- one that is isomorphic to the product of lattices
corresponding to the atoms in $\mc{A}$.  

Given a design $C$, let ${\tr}_C: \mc{S} \rightarrow \mc{S}$ define
the transition function of $C$.  Thus, given a state $s$ of $C$ at
time $t$, the next state of the design at time $t+1$ is given by
${\tr}_C(s)$.  To model the behavior of a design over time, we define
a \emph{sequence} of states as a mapping $\sigma: \mathbb{N}
\rightarrow \mc{S}$, where $\mathbb{N}$ denotes the set of natural
numbers.  A \emph{trajectory} for a design $C$ is a sequence $\sigma$
such that for all $t \in \mathbb{N}$, ${\tr}_C(\sigma(t)) \sqsubseteq
\sigma(t+1)$.  Given two sequences $\sigma_1$ and $\sigma_2$, we abuse
notation and say that $\sigma_1 \sqsubseteq \sigma_2$ iff for every $t
\in \mathbb{N}$, $\sigma_1(t) \sqsubseteq \sigma_2(t)$.

The general \emph{trajectory evaluation logic} of Seger and
Bryant~\cite{SegerBryant95} can be instantiated to words as follows.
A \emph{trajectory formula} is a formula generated by the grammar
\begin{tabular}{lllllllll}
$\varphi$ & ::= & $\mathsf{a}$ is $\mathsf{val}$ & $\mid$ & $\varphi$ and $\varphi$ & $\mid$
                  $P \rightarrow \varphi$ & $\mid$ & $N \varphi$  &
\end{tabular}, where
$\mathsf{a}$ is an atom of $C$, $\mathsf{val}$ is a non-$X$,
non-$\top$ value in the lattice of values for $\mathsf{a}$, and $P$ is
a quantifier-free formula in the theory of bit-vectors.  Formulas like
$P$ in the grammar above are also called \emph{guards} in STE
parlance.

Following Seger et al~\cite{BryantSeger90,SegerJOMABS05}, the
\emph{defining sequence} of a trajectory formula $\psi$ given the
assignment $\phi$, denoted $[\psi]^\phi$, is defined inductively as
follows.  Here, $\mathsf{b}$ denotes an arbitrary $m$-bit atom in
$\mc{A}$ and $t \in \mathbb{N}$.
\begin{itemize}
\item $[\mathsf{a} \text{ is } \mathsf{val}]^\phi(t)(\mathsf{b})
  \isdef \mathsf{val}$ if $t = 0$ and both $\mathsf{a}, \mathsf{b}$
  denote the same $m$-bit atom, and is $\mathbf{X}$ otherwise.
\item $[\psi_1 \text{ and } \psi_2]^\phi(t)(\mathsf{b}) \isdef [\psi_1]^\phi(t)(\mathsf{b}) \;\sqcup_m\; [\psi_2]^\phi(t)(\mathsf{b})$
\item $[P \rightarrow \psi]^\phi(t)(\mathsf{b}) \isdef [\psi]^\phi(t)(\mathsf{b})$ if $\phi \models P$, and is $\mathbf{X}$ otherwise.
\item $[N \psi]^\phi(t)(\mathsf{b}) \isdef
  [\psi]^\phi(t-1)(\mathsf{b})$ if $t \neq 0$, and is $\mathbf{X}$
  otherwise.
\end{itemize}
Similarly, the \emph{defining trajectory} of $\psi$ with respect to
a design $C$, denoted $\lsem \psi \rsem_C^\phi$ can be defined as
follows.
\begin{itemize}
\item $\lsem \psi \rsem_C^\phi(0) \isdef [\psi]^\phi(0)$
\item $\lsem \psi \rsem_C^\phi(t+1) \isdef [\psi]^\phi(t+1) \;\sqcup\;  {\tr}_C(\lsem \psi \rsem_C^\phi(t))$ for every $t \in \mathbb{N}$.
\end{itemize}
In symbolic trajectory evaluation, we are given an antecedent $\ant$
and a consequent $\cons$ in trajectory evaluation logic.  We are also
given a quantifier-free formula $\constr$ in the theory of bit-vectors
with free variables that appear in the guards of $\ant$ and/or
$\cons$.  We wish to determine if for every assignment $\phi$ that
satisfies $\constr$, we have $[\cons]^\phi \sqsubseteq \lsem \ant
\rsem_C^\phi$.


\subsection{Implementation}
We have developed a tool called {\steword} that uses symbolic
simulation with invalid-bit encoding and SMT solving to perform STE.
Each antecedent and consequent tuple has the format $(g,
a, vexpr, start, end)$, where $g$ is a guard, $a$ is the name of an
atom in the design under verification, $vexpr$ is a symbolic
expression over constants and guard variables that specifies the value
of $a$, and $start$ and $end$ denote time points such that $end \ge
start + 1$.

An antecedent tuple $(g, a, vexpr, t_1, t_2)$ specifies that given an
assignment $\phi$ of guard variables, if $\phi \models g$, then atom
$a$ is assigned the value of expression $vexpr$, evaluated on satisfying
assignments of $\phi$,
for all time in $\{t_1, \ldots t_2 - 1\}$.  If, however, $\phi
\not\models g$, atom $a$ is assigned the value $\mathbf{X}$ for all
time in $\{t_1, \ldots t_2-1\}$.  If $a$ is an input atom, the
antecedent tuple effectively specifies how it is driven from time
$t_1$ through $t_2-1$.  Using invalid-bit encoding, the above
semantics is easily implemented by setting ${\inv}(a)$ to $\neg g$ and
${\val}(a)$ to $vexpr$ from time $t_1$ through $t_2-1$.  If $a$ is an
internal atom, the defining trajectory requires us to compute the
$lub$ of the value driven by the circuit on $a$ and the value
specified by the antecedent for $a$, at every time point in $\{t_1,
\ldots t_2-1\}$.  The value driven by the circuit on $a$ at any time
is computed by symbolic simulation using invalid-bit encoding, as
explained in Sections~\ref{subsec:sim-val} and \ref{subsec:sim-inv}.
The value driven by the antecedent can also be invalid-bit encoded, as
described above.  Therefore, the $lub$ can be computed as described in
Section~\ref{subsec:lub}.  If the $lub$ is not $\top$, ${\val}(a)$ and
${\inv}(a)$ can be set to the value and invalid-bit, respectively, of
the $lub$.  In practice, we assume that the $lub$ is not $\top$ and
proceed as above. The conditions under which the $lub$ evaluates to
$\top$ are collected separately, as described below.  The values of
all atoms that are not specified in any antecedent tuple are obtained
by symbolically simulating the circuit using invalid-bit encoding.

If the $lub$ computed above evaluates to $\top$, we must set atom $a$
to an unachievable over-constrained value.  This is called
\emph{antecedent failure} in STE parlance.  In our implementation, we
collect the constraints (condition for case (a) in
Section~\ref{subsec:lub}) under which antecedent failure occurs for
every antecedent tuple in a set {\AntFail}.  Depending on the mode of
verification, we do one of the following:
\begin{itemize}
\item If the disjunction of formulas in {\AntFail} is satisfiable, we
  conclude that there is an assignment of guard variables that leads
  to an antecedent failure.  This can then be viewed as a failed run
  of verification.
\item We may also wish to check if $[\cons]^\phi \sqsubseteq \lsem
  \ant \rsem_C^\phi$ only for assignments $\phi$ that do not satisfy
  any formula in ${\AntFail}$.  In this case, we conjoin the negation
  of every formula in ${\AntFail}$ to obtain a formula, say
  ${\NoAntFail}$, that defines all assignments $\phi$ of interest.
\end{itemize}

A consequent tuple $(g, a, vexpr, t_1, t_2)$ specifies that given an
assignment $\phi$ of guard variables, if $\phi \models g$, then atom
$a$ must have its invalid bit set to ${\false}$ and value set to
$vexpr$, evaluated on satisfying assignments of $\phi$, for all time
in $\{t_1, \ldots t_2 - 1\}$.  If $\phi \not\models g$, a
consequent tuple imposes no requirement on the value of atom $a$.
Suppose that at time $t$, a consequent tuple specifies a guard $g$ and
a value expression $vexpr$ for an atom $a$.  Suppose further that
$({\val}(a), {\inv}(a))$ gives the invalid-bit encoded value of this
atom at time $t$, as obtained from symbolic simulation.  Checking
whether $[\cons]^\phi(t)(a) \sqsubseteq \lsem \ant \rsem_C^\phi(t)(a)$
for all assignments $\phi$ reduces to checking the validity of the
formula $\big(g \rightarrow (\neg{\inv}(a) \wedge (vexpr = {\val}(a)))
\big)$.  Let us call this formula $OK_{a, t}$.  Let $\mc{T}$ denote
the set of all time points specified in all consequent tuples, and let
$\mc{A}$ denote the set of all atoms of the design.  The overall
verification goal then reduces to checking the validity of the formula
$OK \isdef \bigwedge_{t \in \mc{T},\;a \in \mc{A}} OK_{a,t}$.  If we
wish to focus only on assignments $\phi$ that do not cause any
antecedent failure, our verification goal is modified to check the
validity of ${\NoAntFail} \rightarrow OK$.  In our implementation, we
use {\Boolector}~\cite{BrummayerBiere09}, a state-of-the-art solver
for bit-vectors and the extensional theory of arrays, to check the
validity (or satisfiability) of all formulas $OK$ generated by
{\steword}.

\section{Experiments}\label{sec:experiments}
We used {\steword} to verify properties of a set of System-Verilog
word-level benchmark designs.  Bit-level STE tools are often known to
require user-guidance with respect to problem decomposition and
variable ordering (for BDD based tools), when verifying properties of
designs with moderate to wide datapaths.  Similarly, BMC tools need to
introduce a fresh variable for each input in each time frame when the
value of the input is unspecified.  Our benchmarks were intended to
stress bit-level STE tools, and included designs with control and
datapath logic, where the width of the datapath was parameterized.
Our benchmarks were also intended to stress BMC tools by providing
relatively long sequences of inputs that could either be $X$ or a
specified symbolic value, depending on a symbolic condition.  In each
case, we verified properties that were satisfied by the system and
those that were not.  For comparative evaluation, we implemented
word-level bounded model checking as an additional feature of
{\steword} itself.  Below, we first give a brief description of each
design, followed by a discussion of our experiments.

{\bfseries \emph{Design 1:}} Our first design was a three-stage
pipelined circuit that read four pairs of $k$-bit words in each cycle,
computed the absolute difference of each pair, and then added the
absolute differences with a current running sum.  Alternatively, if a
reset signal was asserted, the pipeline stage that stored the sum was
reset to the all-zero value, and the addition of absolute differences
of pairs of inputs started afresh from the next cycle.  In order to
reduce the stage delays in the pipeline, the running sum was stored in
a redundant format and carry-save-adders were used to perform all
additions/subtractions. Only in the final stage was the non-redundant
result computed. In addition, the design made extensive use of clock
gating to reduce its dynamic power consumption -- a characteristic of
most modern designs and that significantly complicates formal
verification.  Because of the non-trivial control and
clock gating, the STE verification required a simple datapath
invariant.  Furthermore, in order to reduce the complexity in
specifying the correctness, we broke down the overall verification
goal into six properties, and verified these properties using several
datapath widths.

{\bfseries \emph{Design 2:}} Our second design was a pipelined serial
multiplier that read two $k$-bit inputs serially from a single $k$-bit
input port, multiplied them and made the result available on a
$2k$-bit wide output port in the cycle after the second input was
read.  The entire multiplication cycle was then re-started afresh.  By
asserting and de-asserting special input flags, the control logic
allowed the circuit to wait indefinitely between reading its first and
second inputs, and also between reading its second input and making
the result available.  We verified several properties of this circuit,
including checking whether the result computed was indeed the product
of two values read from the inputs, whether the inputs and results
were correctly stored in intermediate pipeline stages for various
sequences of asserting and de-asserting of the input flags, etc.  In
each case, we tried the verification runs using different values of
the bit-width $k$.

{\bfseries \emph{Design 3:}} Our third design was an implementation of
the first stage in a typical digital camera pipeline.  The design is
fed the output of a single CCD/CMOS sensor array whose pixels have
different color filters in front of them in a Bayer mosaic pattern
\cite{MalvarLiWeiCutler04}. The design takes these values and performs
a ``de-mosaicing'' of the image, which basically uses a fairly
sophisticated interpolation technique (including edge detection) to
estimate the missing color values.  The challenge here was not only
verifying the computation, which entailed adding a fairly large number
of scaled inputs, but also verifying that the correct pixel values
were used. In fact, most non-STE based formal verification engines will
encounter difficulty with this design since the final result depends on
several hundreds of $8$-bit quantities.

{\bfseries \emph{Design 4:}} Our fourth design is a more general
version of Design 3, that takes as input stream of values from a
single sensor with a mosaic filter having alternating colors, and
produces an interpolated red, green and blue stream as output.  Here,
we verify $36$ different locations on the screen, which translates to
$36$ different locations in the input stream. Analyzing this example
with BMC requires providing new inputs every cycle for over $200$
cycles, leading to a blow-up in the number of variables used.

For each benchmark design, we experimented with a bug-free version,
and with several buggy versions.  For bit-level verification, we used
both a BDD-based STE tool~\cite{SegerJOMABS05} and propositional SAT
based STE tool~\cite{RoordaClaessen05}; specifically, the tool
{\forte} was used for bit-level STE.  We also ran word-level BMC to
verify the same properties.

In all our benchmarks, we found that {\forte} and {\steword}
successfully verified the properties within a few seconds when the
bitwidth was small ($8$ bits).  However, the running time of {\forte}
increased significantly with increasing bit-width, and for bit-widths
of $16$ and above, {\forte} could not verify the properties without
serious user intervention.  In contrast, {\steword} required
practically the same time to verify properties of circuits with wide
datapaths, as was needed to verify properties of the same circuits
with narrower datapaths, and required no user intervention.  In
fact, the word-level SMT constraints generated for a circuit with a
narrow datapath are almost identical to those generated for a circuit
with a wider datapath, except for the bit-widths of atoms.  This is
not surprising, since once atomization is done, symbolic simulation is
agnostic to the widths of various atoms.  An advanced SMT solver like
Boolector is often able to exploit the word-level structure of the
final set of constraints and solve it without resorting to
bit-blasting.

The BMC experiments involved adding a fresh variable in each time
frame when the value of an input was not specified or conditionally
specified.  This resulted in a significant blow-up in the number of
additional variables, especially when we have long sequences of
conditionally driven inputs.  This in turn adversely affected
SMT-solving time, causing BMC to timeout in some cases.

To illustrate how the verification effort with {\steword} compared
with the effort required to verify the same property with a bit-level
BDD- or SAT-based STE tool, and with word-level BMC, we present a
sampling of our observations in Table I, where no user intervention
was allowed for any tool.  Here ``-'' indicates more than $2$ hours
of running time, and all times are on an Intel Xeon 3GHz CPU, using
a single core.  In the column labeled ``Benchmark'',
Design$i$-P$j$ corresponds to verifying property $j$ (from a list of
properties) on Design $i$.  The column labeled ``Word-level latches
(\# bits)'' gives the number of word-level latches and the total
number of bits in those latches for a given benchmark.  The column
labeled ``Cycles of Simulation'' gives the total number of time-frames
for which STE and BMC was run.  The column labeled ``Atom Size
(largest)'' gives the largest size of an atom after our atomization
step.  Clearly, atomization did not bit-blast all words,
allowing us to reason at the granularity of multi-bit atoms in {\steword}.

\begin{table}
\scriptsize
\begin{center}
\begin{tabular}{|c|c|c|c|c|c|c|}
\hline
Benchmark & {\steword} & {\forte}  & BMC & Word-level latches & Cycles of  & Atom Size \\
        &            & (BDD and SAT) &     & (\# bits)     & Simulation & (largest) \\
\hline
Design1-P1 & 2.38s                   & -                             & 3.71s & 14 latches & 12 & 31 \\
(32 bits)  &         & -                                   &  & (235 bits wide) & & \\
\hline
Design1-P1 & 2.77s                   & -                                   & 4.53s & 14 latches & 12 & 64 \\
(64 bits)  &         & -                                   &  & (463 bits wide) & & \\
\hline
Design2-P2 & 1.56s                   & -                             & 1.50s & 4 latches & 6 & 32\\
(16 bits)  &         & -                                   &  & (96 bits wide) & & \\
\hline
Design2-P2 & 1.65s                   & -                                   & 1.52s & 4 latches & 6 & 64\\
(32 bits)  &         & -                                   &  & (128 bits wide) & & \\
\hline
Design3-P3 & 24.06s                  & -                                   & - & 54 latches & 124 & 16 \\
(16 bits)  &         & -                                   &  & (787 bits wide) & & \\
\hline
Design4-P4 & 56.80s                  & -                                   & - & 54 latches & 260 & 16 \\
(16 bits)  &         & -                                   &  & (787 bits wide) & & \\
\hline
Design4-P4 & 55.65s                  & -                                   & - & 54 latches & 260 & 32 \\
(32 bits)  &         & -                                   &  & (1555 bits wide) & & \\
\hline
\end{tabular}
\end{center}
\caption{Comparing verification effort (time) with {\steword}, {\forte} and BMC}
\vspace*{-0.2in}
\label{compare-table}
\end{table}
Our experiments indicate that when a property is not satisfied by a
circuit, Boolector finds a counterexample quickly due to powerful
search heuristics implemented in modern SMT solvers.  BDD-based
bit-level STE engines are, however, likely to suffer from BDD size
explosion in such cases, especially when the bit-widths are large.  In
cases where there are long sequences of conditionally driven inputs
(e.g., design 4) BMC performs worse compared to {\steword}, presumably
beacause of the added complexity of solving constraints with
significantly larger number of variables.  In other cases, the
performance of BMC is comparable to that of {\steword}.  An important
observation is that the abstractions introduced by atomization and by
approximations of invalid-bit expressions do not cause {\steword} to
produce conservative results in any of our experiments.  Thus,
{\steword} strikes a good balance between accuracy and performance.
Another interesting observation is that for correct designs and
properties, SMT solvers (all we tried) sometimes fail to verify the
correctness (by proving unsatisfiability of a formula).  This points
to the need for further developments in SMT solving, particularly for
proving unsatisfiability of complex formulas.  Overall, our
experiments, though limited, show that word-level STE can be
beneficial compared to both bit-level STE and word-level BMC in
real-life verification problems.

We are currently unable to make the binaries or source of {\steword}
publicly available due to a part of the code being proprietary.  A
web-based interface to {\steword}, along with a usage document and the
benchmarks reported in this paper, is available at
http://www.cfdvs.iitb.ac.in/WSTE/

\section{Conclusion}\label{sec:conclusion}
Increasing the level of abstraction from bits to words is a promising
approach to scaling STE to large designs with wide datapaths.  In this
paper, we proposed a methodology and presented a tool to achieve this
automatically.  Our approach lends itself to a
counterexample guided abstraction refinement (CEGAR) framework, where
refinement corresponds to reducing the conservativeness in invalid-bit
expressions, and to splitting existing atoms into finer bit-slices.
We intend to build a CEGAR-style word-level STE tool as part of future
work.

\subsubsection{Acknowledgements.} We thank Taly Hocherman and Dan Jacobi for
their help in designing a System-Verilog symbolic simulator.  We thank
Ashutosh Kulkarni and Soumyajit Dey for their help in implementing and
debugging {\steword}.  
\bibliographystyle{plain} 

\end{document}